\documentclass[prl,aps, amsfonts, 11pt, nofootinbib, notitlepage]{revtex4-1}
\usepackage[font=small,format=plain,labelfont=bf,up,textfont=normal,up,justification=justified,singlelinecheck=false]{caption}

\usepackage[paperwidth=210mm,paperheight=297mm,centering,hmargin=1.90cm,vmargin=2.05cm]{geometry}
\usepackage{caption}
\usepackage{color}

\usepackage{subfigure}
\usepackage{hyperref}
\usepackage{epsfig}
\usepackage{amssymb}
\usepackage{amsthm}
\usepackage{amsmath}
\usepackage{amsmath}

\newcommand{\beq}{\begin{equation}}
\newcommand{\eeq}{\end{equation}}
\newcommand{\bea}{\begin{eqnarray}}
\newcommand{\eea}{\end{eqnarray}}

\newcommand{\nn}{\nonumber}

\newcommand{\benn}{\begin{displaymath}}
\newcommand{\eenn}{\end{displaymath}}

\begin{document}
\begin{flushright}
{NT@UW-12-18}
\end{flushright}

\title{{Three-particle scattering amplitudes from a finite volume formalism}}

\author{Ra\'ul A. Brice\~no\footnote{{\tt briceno@uw.edu}} and Zohreh Davoudi
\footnote{{\tt davoudi@uw.edu}}}
\affiliation{Department of Physics, University of Washington\\
Box 351560, Seattle, WA 98195, USA}

\begin{abstract}
We present a quantization condition for the spectrum of a system composed of three identical bosons in a finite volume with periodic boundary conditions. This condition gives a relation between the finite volume spectrum and infinite volume scattering amplitudes. The quantization condition presented is an integral equation that in general must be solved numerically. However, for systems with an attractive two-body force that supports a two-body bound-state, a \emph{diboson}, and for energies below the diboson breakup, the quantization condition reduces to the well-known L\"uscher formula with exponential corrections in volume that scale with the diboson binding momentum. To accurately determine infinite volume phase shifts, it is necessary to extrapolate the phase shifts obtained from the L\"uscher formula for the boson-diboson system to the infinite volume limit. For energies above the breakup threshold, or for systems with no two-body bound-state (with only scattering states and resonances) the L\"uscher formula gets power-law volume corrections and consequently fails to describe the three-particle system. These corrections are nonperturbatively included in the quantization condition presented.

 \end{abstract}
\maketitle
\section{Introduction \label{intro}}

Quantum chromodynamics (QCD) as the underlying theory of strong interactions is responsible for a wide range of phenomena in nuclear and particle physics. Non-perturbative numerical evaluations of the QCD path integral through lattice QCD (LQCD) have started to bridge the gap between nuclear physics and the underlying theory of QCD. With ever-increasing computational resources and ever-decreasing uncertainties in LQCD calculations of various quantities, it is reasonable to expect few-body nuclear reaction cross-sections to be reliably evaluated in upcoming years. This program will allow for the determination of nuclear few-body forces in a model-independent fashion which will consequently lead to a reduction of the systematic uncertainties of many-body calculations. Reaching that point, however, requires developing the formalism to relate the physical observables of interest to the LQCD correlation functions that are evaluated in a finite Euclidean spacetime.

For two-body systems below inelastic thresholds, the energy spectrum in a finite cubic volume with periodic boundary conditions fully determines the infinite volume scattering phase shifts, up to exponential corrections in the volume. These corrections are negligible provided that interactions are fully localized inside the volume. This idea, which is due to L\"uscher \cite{luscher1, luscher2}, has been extensively implemented in obtaining scattering parameters and binding energies of meson-meson, meson-baryon and baryon-baryon systems (see Refs. \cite{Beane:2010hg, Beane:2011xf, Beane:2011iw, Beane:2011sc, Inoue:2010es, Inoue:2011ai, Dudek:2012gj} for some recent works in this direction). This approach has also motivated studies of several hadronic resonances via LQCD as reviewed in Ref. \cite{Mohler:2012nh}; in particular the two-body coupled-channel generalization of the L\"uscher formalism \cite{He:2005ey, Lage:2009zv, Bernard:2010fp, Hansen:2012tf, Briceno:2012yi} allows one to access physics beyond inelastic thresholds where new two-particle channels open up. Going beyond two particles in this finite volume (FV) formalism, however, has been a major obstacle in the field. Recently several efforts have been initiated to address this problem via different approaches \cite{Kreuzer:2008bi, Kreuzer:2009jp, Kreuzer:2010ti, Kreuzer:2012sr, Polejaeva:2012ut, Bour:2011ef, Bour:2012hn, Guo:2012hv}; aiming to make a connection between the elements of the S-matrix and the spectrum of the three-particle system in a finite volume. These benchmark theoretical calculations give one confidence that the infinite volume elements of the S-matrix can in fact be determined from the discretized spectrum in the finite volume even above the inelastic threshold, and has motivated us to further explore this connection using a non-relativistic field theoretical approach.

In this article, we derive the quantization condition (QC) for the spectrum of a system composed of three identical bosons in a finite volume with periodic boundary conditions. The quantization condition gives a relation between the FV spectrum and infinite volume scattering amplitudes, Eqs. (\ref{QC}), (\ref{STMeq2}). This quantization condition in general must be solved numerically, since the relation between the FV energy eigenvalues and three-particle scattering amplitudes is not algebraic. We pay close attention to systems with an attractive two-body force that allows for a two-body bound-state, a \emph{diboson}, and energies below the diboson breakup. For these theories in this energy regime, the quantization condition reduces to the well-known L\"uscher formula with exponential corrections in volume with a length scale that is dictated by the inverse diboson binding momentum. In other words, the boson-diboson scattering phase shifts can be obtained from the three-particle spectrum using the following relation
\begin{eqnarray}
\label{dbQC}
{q}^*_{0}\cot\delta_{Bd} &=&4\pi \ c^P_{00}({q}_{0}^{*})+\eta\frac{e^{-\gamma_d L}}{L} \ ,
\end{eqnarray}
where ${q}_{0}^{*}=\sqrt{\frac{4}{3}\left(mE^*+\bar{q}_{0}^{*2}\right)}$ is the momentum of the boson and diboson in the center of mass (CM) frame of boson-diboson system with $\bar{q}_{0}^{*}$ being the relative momentum of the two bosons in the diboson in the CM frame of the diboson, $m$ is the mass of the three identical particles, $E^*$ is the CM energy, $\gamma_d$ is the binding momentum of the diboson in the infinite volume limit, $\delta_{Bd}$ is the scattering phase shift of the boson-diboson system, $L$ is the spatial extent of the cubic volume, and $\eta$ is an unknown coefficient that must be fitted when extrapolating results to the infinite volume. Given that the diboson is bound, $\bar{q}_{0}^{*2}<0$ and $\bar{q}_{0}^{*2}\rightarrow -\gamma_d^2$ as the volume goes to infinity. $c^{{P}}_{lm}$ is a kinematic function that is given by
\begin{eqnarray}
\label{clm}
c^{{P}}_{lm}(x)=\left[\frac{1}{L^3}\sum_{\textbf{k}}-\mathcal{P}\int\frac{d^3\mathbf{k}}{(2\pi)^3}\right]\frac{\sqrt{4\pi}Y_{lm}(\hat{k}^*)}{{k}^{*2}-x} \ ,
\end{eqnarray}
where $\mathcal{P}$ denotes the principal value of the integral, and for non-relativistic particles $\mathbf{k}^*=\mathbf{k}-\alpha\mathbf{P}$. $\mathbf{P}$ is the total momentum of the boson-diboson system, and $\alpha=\frac{m_1}{m_1+m_2}$ \cite{Bour:2011ef}; so for a diboson that is twice as massive as the boson $\alpha=\frac{1}{3}$. In addition to exponential corrections that are governed by the size of the bound-state wavefunction, there are other exponential volume corrections to the above L\"uscher relation that are arising from the the off-shell states of the 2+1 system. These corrections however are subleading compared to the exponential corrections denoted in Eq. (\ref{dbQC}), and will be discussed in Sec. \ref{recoverL} in more details.

In order to reliably use such analytical formula one must necessarily be in the regime where $\gamma_d L$ is at least $4$ so that the infinite volume phase shifts of the bound-state particle scattering can be obtained with a few percent uncertainty. This is an important distinction compared to the the two-body problem where the dominant finite volume corrections to the L\"uscher formula scale like $\sim e^{-m_\pi L}$ where $m_{\pi}$ denotes the mass of the pion. So although volumes of the order of $6~\rm{fm}$ or greater would reliably recover, for example, $\pi\pi$ scattering phase shifts at the physical pion mass, in order to accurately recover phase shifts for deuteron-neutron scattering one would naively need $L\gtrsim17~\rm{fm}$. But presumably upon quantifying the coefficients of these exponentials, linear combinations of these exponential corrections can be formed for different boost momenta of the three-particle system so that to cancel out the leading exponential corrections to the above quantization condition, and therefore reduce the size of the volumes needed for a reliable determination of the phase shifts to $L\gtrsim12~\rm{fm}$.\footnote{For a discussion of the improvement of the volume dependence of deuteron binding energy in LQCD calculations see Ref. \cite{Davoudi}.} Note that the next to leading exponential correction due to the size of the bound-state scale as $e^{-\sqrt{2}\gamma_d L}/L$. The quantization condition shown in Section \ref{3body} demonstrates that for energies above the diboson breakup Eq. (\ref{dbQC}) gets power-law corrections associated with new possible states that can go on-shell and the quantization condition must be solved numerically.

Studying few-body systems in a finite volume is not only pertinent for the determination of nuclear reactions cross sections, but will also have an impact in our understanding of excited states and resonances of QCD. This has recently motivated Guo \emph{et al.}~\cite{Guo:2012hv} to explore the three-body problem in a finite volume. The formalism developed in this article, however, clearly illustrates that the quantization condition for energy eigenvalues of the three-particle system in a finite volume in the isobar approximation \cite{Goradia:1975ec, Ascoli:1975mn, Hwa:1963, Aaron:1973ca} that is derived in Ref.~\cite{Guo:2012hv}, fails to describe the physics of the three-particle processes these authors have been interested in, for several reasons that are summarized in subsection \ref{isobarf}. Basically although the quantization condition of Ref.~\cite{Guo:2012hv} takes a simple algebraic form, there are power-law corrections in volume for resonant two-body sub-systems (which has been the motivation of Ref. \cite{Guo:2012hv} for studying the three-body systems) that invalidates the use of their result. These corrections arise from the fact that the two-particle sub-system is not compact but rather an extended object that, that regardless of its resonant nature, can sample the boundaries of the volume. These power-law corrections are nonperturbatively included in the quantization condition presented, Eqs. (\ref{QC}). The quantization conditions of Ref. \cite{Guo:2012hv} also do not incorporate the possibility of new on-shell three-particle channels that should turn the quantization conditions to a coupled-channels form.

In addition, we comment on another source of systematic uncertainty in previous finite volume three-body calculations \cite{Kreuzer:2008bi, Kreuzer:2009jp, Kreuzer:2010ti, Kreuzer:2012sr} that has not been fully addressed. This systematic uncertainty is due to the reduced symmetry of the calculations in a finite cubic volume with periodic boundary conditions. In an S-wave 2+1 scattering where the two-body sub-system is also in the S-wave, the nearest partial-wave mixing, which is familiar from the L\"uscher formalism, occurs with the D-wave two-particle state for the identical particles \cite{movingframe}. This is due to the fact that in the CM frame of the three-particle system, the two-particle sub-system is necessarily boosted. As a result, the spherically symmetric S-wave dimer field which has been proven to be well suited for studying the three-particle system in the infinite volume \cite{Skorniakov, Bedaque:1997qi,Bedaque:1998mb}, does not incorporate this additional feature of two-particle wave-function in a finite cubic volume. {In other words, the D-wave partial wave mixing has been unaccounted for in previous numerical calculations by Kreuzer \emph{et al.} \cite{Kreuzer:2008bi, Kreuzer:2009jp, Kreuzer:2010ti, Kreuzer:2012sr}, which is the leading systematic error of their results. }

The rest of this article is organized as follows. In Sec. \ref{3body}, after introducing the dimer formalism, we review the connection between the finite volume spectrum of the boosted system and the two-body scattering phase shifts. The advantages as well as limitations of the dimer formalism for this particular calculation are discussed. By means of the dimer formalism, the energy spectrum of three identical bosons with two-body and three-body interactions in the finite volume is determined from the poles of the finite volume three-body correlation function. The features of this quantization condition for the energy eigenvalues of the system are discussed. In particular, it is demonstrated that this condition gives an integral equation that relates the spectrum and the elements of the S-matrix. In Sec. \ref{recoverL}, the L\"uscher formula for the elastic scattering of a bound-state and a particle is deduced from the energy quantization condition. The systematic corrections due to the size of the composite object as well as those due to the presence of off-shell states are identified. We devote subsection \ref{isobarf} to comparing the result of Ref. \cite{Guo:2012hv} to the result presented in this article, and will comment on its validity in studying three-body systems in the isobar model. We conclude in Sec. \ref{summ}, and briefly discuss the strategies that a practitioner needs to follow in order to obtain the scattering quantities of the three-particle system from a LQCD calculation of the spectrum.
 
\color{black}

\section{Three Particles in a Finite Volume: Quantization Condition \label{3body}}
In studying a three-body problem, it is customary to divide the system of three particles to a system of two particles interacting in a given partial-wave $J_{d}$, and a third particle, called the spectator, which interacts with the two-body system with angular momentum $J_{Bd}$. In particular, the diagrammatic representation of few-body scattering amplitudes is greatly simplified when using an auxiliary S-wave \emph{dimer} field \cite{pionless2, pionless3} which non-perturbatively sums all $2 \rightarrow 2$ diagrams. 

Consider three bosons with mass $m$ and total energy and momentum $(E,\mathbf{P})$ in the lab frame. The total CM energy of the three-particle system, $E^*$ is then given by $E^{*}=E-\frac{{P}^2}{6m}$. Also the relative momentum of the spectator boson and the dimer in the CM frame of three particles, $\mathbf{q}^{*}$, is related to the momentum of the spectator boson in the lab frame, $\mathbf{q}$ by ${\mathbf{q}}^{*}={\mathbf{q}}-\frac{\textbf{P}}3$. The total CM energy of boson-dimer system can be written as $E^*=\overline{q}^{*2}/m+3{q}^{*2}/4m$, where $\overline{\mathbf{q}}^{*}$ is the relative momentum of the two bosons inside the dimer. 

\begin{figure}[t]
\begin{center}
\subfigure[]{
\includegraphics[scale=0.45]{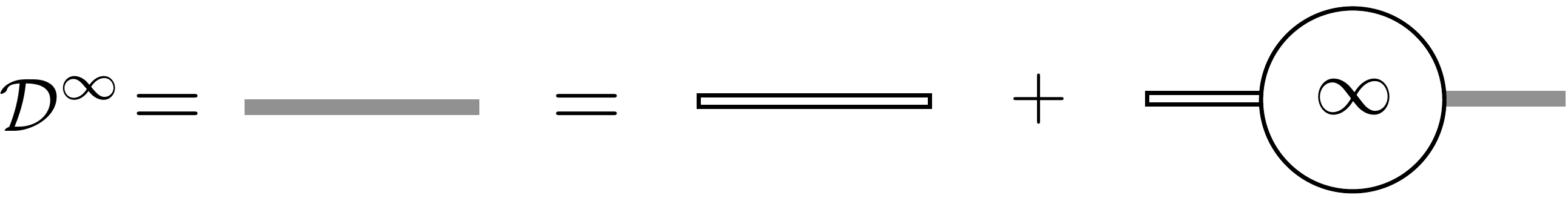}}
\subfigure[]{
\includegraphics[scale=0.45]{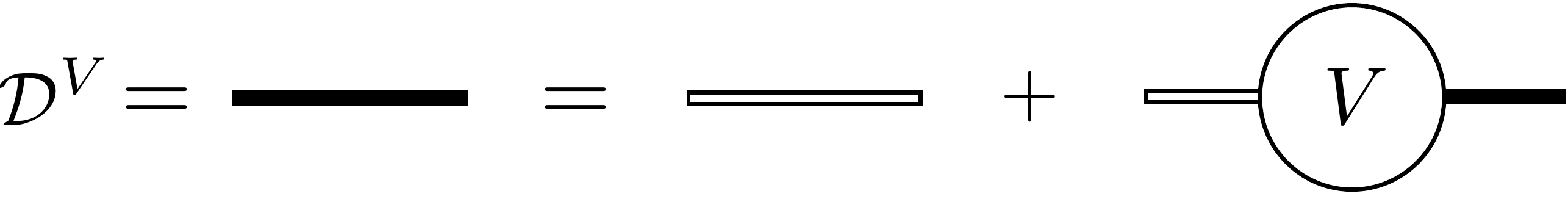}}
\caption{a) Diagrammatic equation satisfied by the full dimer propagator in a) infinite volume and b) finite volume. The grey (black) band represents the full infinite (finite) volume propagator while the double lines represent the bare propagator.}\label{fig:dimer}
\end{center}
\end{figure}

One can write the full dimer propagator in the infinite volume by summing up all the $2 \rightarrow 2$ interactions as depicted in Fig. (\ref{fig:dimer}, a). By requiring the infinite volume two-body scattering amplitude to be recovered from the full dimer propagator, one obtains
\begin{eqnarray}
\label{dimerprop}
i\mathcal{D}^{\infty}(E_2,\mathbf{q}^*)=
\frac{-imr/2}{\overline{q}^{*}\cot{\delta_d}-i\overline{q}^{*}+i\epsilon} \ ,
\end{eqnarray}
where $E_2=E^*-q^{*2}/2m$ is the total energy of the dimer. $\delta_d$ denotes the S-wave scattering phase shift of the two-boson system, and $r$ is its effective range. Similarly one can work out the finite volume case, Fig. (\ref{fig:dimer}, b), where due to the periodic boundary condition, the lab frame momenta are discretized, $\mathbf{q}=\frac{2 \pi}{L} \mathbf{n}$ where $\mathbf{n}$ is a vector with integer components, and the loop integrations are replaced by the corresponding sums, $\int\frac{d^3q}{(2\pi)^3}\rightarrow \frac{1}{L^3}\sum_{\mathbf{q}}$. Then it is straightforward to show that the full finite volume dimer propagator can be written as
\begin{eqnarray}
\label{dimerprop}
i\mathcal{D}^{V}(E_2,\mathbf{q}^*)=
\frac{-imr/2}{\overline{q}^{*} \cot{\delta_d}-4\pi \ c^{{q}^*}_{00}(\overline{q}^{*2}{+i\epsilon}) +i\epsilon} \ ,
\end{eqnarray}
where the kinematic function $c^{{q}^*}_{lm}$ is defined in Eq. (\ref{clm}).
For the case of a dimer composed of identical bosons, one therefore has $\mathbf{k}^*=\mathbf{k}-\frac{\mathbf{q^*}}{2}$.%

The spectrum in a finite volume can be obtained from the poles of the two-particle propagator or equivalently from the poles of the finite volume dimer, Eq. (\ref{clm}),
\begin{eqnarray} 
\label{2QCV}
\overline{q}^{*}_{\kappa} \cot{\delta_d}=4\pi \ c^{{q}^*}_{00}(\overline{q}_{\kappa}^{*2}) \ ,
\end{eqnarray}
where $\overline{q}_{\kappa}$ is the $\kappa^{th}$ solution to the quantization condition for a boosted two particle system \cite{movingframe, sharpe1}. As will be discussed in great details, these poles play an important role in the three-body sector and will be referred to as \emph{L\"uscher} poles. Note that this result is equivalent to the non-relativistic limit of the result obtained in Refs. \cite{movingframe, sharpe1, Christ:2005gi} for the boosted systems of particles with identical masses, and for that of systems with unequal masses~\cite{Davoudi, Fu:2011xz, Leskovec:2012gb}.

The simplification due to the use of the dimer field comes at the cost of truncating the precision of the effective range expansion of the two-body sector at next to leading order (NLO). This, however, can be shown to be systematically improved by including higher order terms in the effective range expansion in constructing such a dimer field.
The other systematic error associated with introducing such an auxiliary field in the finite volume arises from the fact that the spectrum that is obtained by looking at the poles of the dimer propagator corresponds to two bosons in an S-wave, and does not incorporate the partial-wave mixing. This partial-wave mixing is expected since the eigenstates of Hamiltonian in a finite cubic volume do not respect the full symmetries of the rotational group, and are instead identified by the irreducible representations (irreps) of the Cubic (octahedral) group. The irreps of the cubic group are in general reducible under the irreps of the full rotation group. So, for example, the ground-state of two bosons in a finite cubic volume is an eigenstate of the $A_1$ irreducible representation of the cubic group, which in the CM reference frame has overlap with not only $J=0$ but also $J=4,6,8,\ldots$ \cite{luscher1, luscher2}. For two identical particles moving with a non-zero total momentum, the symmetry is further reduced and the ground state has overlap with $J=0,2,4,6,8,\ldots$ \cite{movingframe, sharpe1, Christ:2005gi}. Although, this mixing between higher partial-waves is explicitly manifested in the pole structure of the two-particle propagator \cite{luscher1, luscher2, movingframe, sharpe1, Christ:2005gi}; it is not present in the dimer formalism. Nevertheless, this technique has proven to be advantageous for studying three-body physics in both infinite and finite volumes \cite{Bedaque:1997qi, Bedaque:1998mb, Bedaque:1998kg, Bedaque:1998km, Gabbiani:1999yv, Bedaque:1999vb, Bedaque:1999ve, Bedaque:2000ft, Kreuzer:2008bi, Kreuzer:2009jp, Kreuzer:2010ti, Kreuzer:2012sr}, and as a first step toward solving the three-body problem in the finite volume, it is reasonable to start with this dimer field. One needs however to keep in mind the systematics of the calculation as presented.

\begin{figure}[t!]
\begin{center}
\subfigure[]{
\includegraphics[scale=0.475]{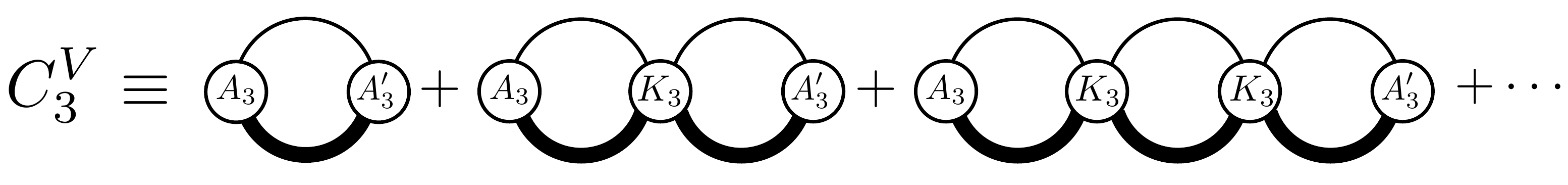}}
\subfigure[]{
\includegraphics[scale=0.35]{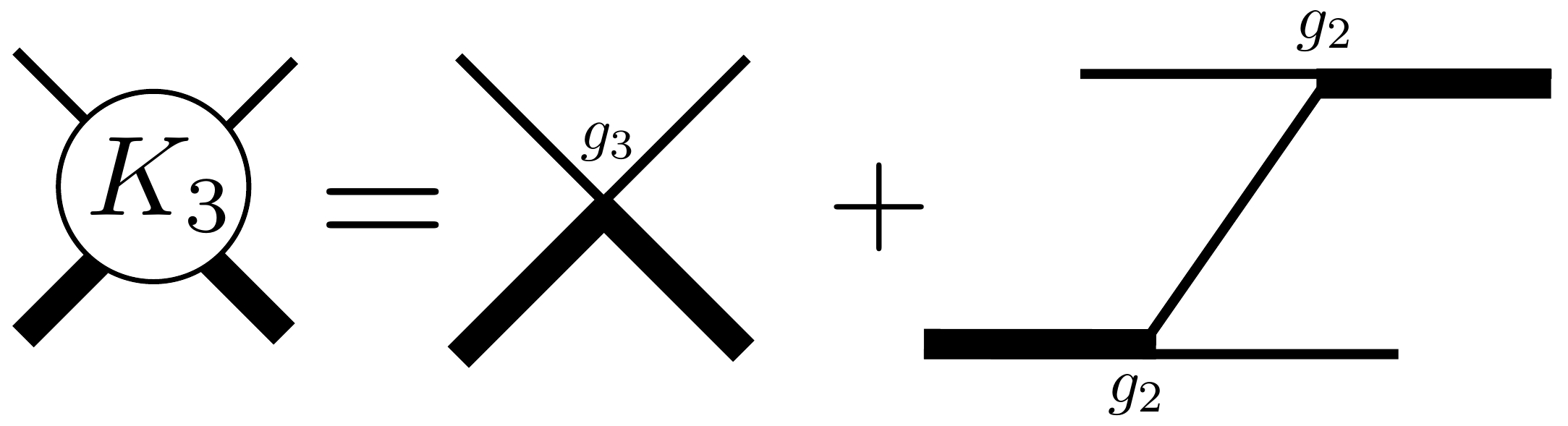}}
\caption{\label{fig:corrfunc} a) Diagrammatic expansion of the three-body correlation function $C_3^V$ in the finite volume. $K_3$ denotes the three-body Bethe-Salpeter kernel and $A_3$ $(A'_3)$ is the overlap the creation (annihilation) dimer-boson interpolating operator has with the initial (final) state with total energy $E$ and total momentum $\mathbf{P}$, b) The effective three-body Bether-Salpeter kernel, $K_3$, is composed of a three-body contact interaction as well as two-body contact interactions via the exchange of a single boson.}
\end{center}
\end{figure}

In order to determine the energy eigenvalues of the three-particle system in a finite volume, one can solve for the poles of the three-particle correlation function as depicted in Fig. (\ref{fig:corrfunc}, a). Algebraically,
\begin{eqnarray}
\label{corr3}
C^V_3(E,\textbf{P})&=&
\frac{1}{L^3}\sum_{\textbf{q}_1}
A_3\left(\mathbf{q}_1\right){i\mathcal{D}^V(E-\frac{{q}^2_1}{2m},|\textbf{P}-\textbf{q}_1|)} \nn\\
&& \times \left[1+\sum_{n'=2}^\infty  \prod_{n=2}^{n'} \left(\frac{1}{L^3} \sum_{\textbf{q}_n}
iK_3(\textbf{q}_{n-1},\textbf{q}_{n};\mathbf{P},E){i\mathcal{D}^V(E-\frac{{q}^2_n}{2m},|\textbf{P}-\textbf{q}_n|)}\right)A'_3\left(\mathbf{q}_{n'}\right)\right] \ ,
\end{eqnarray}
where $A'_3$ $(A_3)$ is the overlap the annihilation (creation) dimer-boson interpolating operator, $\sigma_3$ $(\sigma_3^\dag)$, has with the initial (final) state with total energy $E$ and total momentum $\mathbf{P}$. Note that we have suppressed the total energy and momentum dependence of the overlap factors in our notation. The interactions between three bosons are incorporated in an effective three-body Bethe-Salpeter Kernel, $K_3$, Fig. (\ref{fig:corrfunc}, b),
\begin{eqnarray}
iK_3(\textbf{p},\textbf{k};{\mathbf{P}},E)&\equiv&-{ig_3}-\frac{i{g_2^2}}{E-\frac{\textbf{p}^2}{2m}-\frac{\textbf{k}^2}{2m}-\frac{(\textbf{P}-\textbf{p}-\textbf{k})^2}{2m}+i\epsilon} \ ,
\label{Kernel}
\end{eqnarray}
where the incoming (outgoing) boson has momentum $\mathbf{p}$ $(\mathbf{k})$ and the incoming (outgoing) dimer has momentum $\mathbf{P}-\mathbf{p}$ $(\mathbf{P}-\mathbf{k})$, and $(E,\mathbf{P})$ denote the total energy and momentum of the three-particle system as before. The first term in the Kernel, Eq. (\ref{Kernel}), is the three-body contact interaction, while the second term describes the interaction of three particles via exchange of an intermediate particle through two-body contact interactions.

The finite volume contribution to the first term in the expansion of the three-body correlation function, Eq. (\ref{corr3}), can be evaluated easily using the Poisson resummation formula and the kinematic relations between the CM and lab frame momenta as presented earlier,
\begin{eqnarray}
\label{firstCorr}
\left\{\frac{1}{L^3}\sum_{\textbf{q}_1}-\int\frac{d^3\textbf{q}_1}{(2\pi)^3}\right\}
A_3\left(\mathbf{q}_1\right){i\mathcal{D}^V(E-\frac{{q}^2_1}{2m},|\textbf{P}-\textbf{q}_1|)}A'_3\left(\mathbf{q}_1\right) =
\sum^{N_{E^*}}_{\kappa}A_3({q}_{\kappa}^{*}) \ i \delta\tilde{\mathcal{G}}^V_\kappa({q}_{\kappa}^{*})A'_3({q}_{\kappa}^{*}) \  .
\end{eqnarray}
While $A_3$ and $A'_3$ in the LHS of Eq. (\ref{firstCorr}) are functions of the relative coordinate $\mathbf{q}_1$, they are represented as vectors in the space of the boson-dimer angular momentum, $J_{Bd}$, in the RHS, and are evaluated at the poles of the dimer propagator, ${q}^{*}_{\kappa}$, and the sum runs over these poles. $\delta \tilde{\mathcal{G}}^{V}_{\kappa}$ is a matrix in the same angular momentum basis whose elements are defined by 
\begin{eqnarray}
\label{gtilde}
(\delta\tilde{\mathcal{G}}^V_\kappa)_{l_1m_1,l_2m_2}&\equiv&
\frac{R^V_\kappa}{m}(\delta\mathcal{G}^V_\kappa)_{l_1m_1,l_2m_2} \ ,
\end{eqnarray}
with
\begin{eqnarray}
\label{dgv}
(\delta{\mathcal{G}}^V_\kappa)_{l_1m_1,l_2m_2}&=&i\frac{m~{q}_{\kappa}^{*}}{3\pi} \left(\delta_{l_1,l_2}\delta_{m_1,m_2}+i\sum_{lm}\frac{\sqrt{4\pi}}{{q}_{\kappa}^{*}}c^{{P}}_{lm}({q}_{\kappa}^{*})\int d\Omega~Y^*_{l_1,m_1}Y^*_{l,m}Y_{l_2,m_2}\right) \ .
\end{eqnarray}
The kinematic function $c^{{P}}_{lm}$ is defined in Eq. (\ref{clm}) with $\alpha=\frac{1}{3}$, since the dimer is twice as massive as the boson. The on-shell CM momentum of the boson-dimer system, ${q}^{*}_{\kappa}$, is defined by $\kappa^{th}$ pole of the FV dimer propagator, $\overline{q}^{*2}_{\kappa}=mE^*-\frac{3}{4}{q}^{*2}_{\kappa}$, and $R^{V}_{\kappa}$ is its residue at the $\kappa^{th}$ pole. Explicitly,
\begin{eqnarray}
\label{residue}
\lim_{\overline{q}^{*2}\rightarrow {\overline{q}}_{\kappa}^{*2}}i\mathcal{D}^V(E-\frac{{q}^2}{2m},|\textbf{P}-\textbf{q}|)
\approx \frac{iR^V_\kappa}{\overline{q}^{*2}-{\overline{q}}_{\kappa}^{*2}+i\epsilon}
=
-\frac{4}{3}\frac{iR^V_\kappa}{{q}^{*2}-{q}_{\kappa}^{*2}-i\epsilon} \ ,
\end{eqnarray}
where
\begin{eqnarray}
\label{residuedef}
R^{V}_{\kappa}=-\frac{mr}{2} \left[\left. \frac{\partial}{\partial \overline{q}^{*2}}\left(\overline{q}^{*}\cot\delta_d-{4\pi \ {c^{|\mathbf{P}-\mathbf{q}|}_{00}\left(\overline{q}^{*2}\right)}}\right)\right|_{\overline{q}^{*2}={\overline{q}}_{\kappa}^{*2}}\right]^{-1} \ .
\end{eqnarray}
Note that the poles of the FV dimer propagator correspond to the energy eigenvalues of the boosted two-particle system in the finite volume, Eq. (\ref{2QCV}), i.e. the L\"uscher poles.

\begin{figure}[t]
\begin{center}
\includegraphics[scale=0.375]{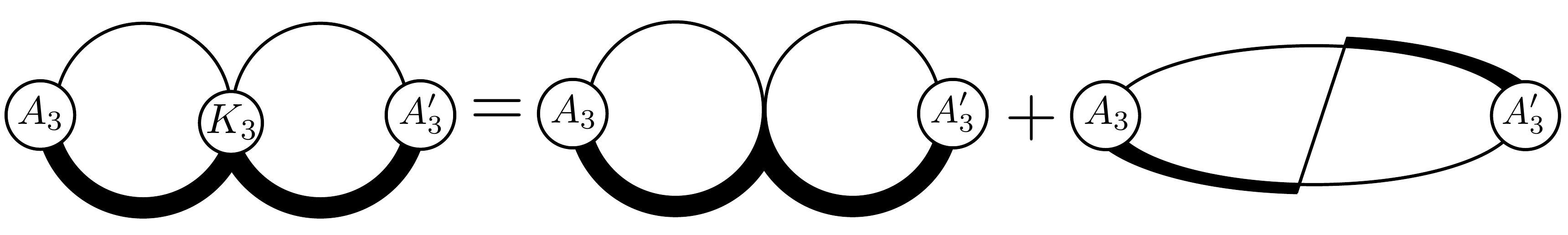}
\caption{\label{fig:twoloop} The NLO contribution to the three-body correlation function with one insertion of the three-body kernel.}
\end{center}
\end{figure}

Equation (\ref{firstCorr}) reflects the fact that, unlike the two-body case, a single on-shell condition does not simultaneously fix the relative momenta of the boson-dimer pair as well as that of the bosons inside the dimer. What is important to observe here is that for a given CM energy $E^*$, there is only a finite number of ``channels" $N_{E^*}$ that can go on-shell, each being identified by a particular configuration of the boson-dimer relative momentum and the relative momentum of the two bosons (of the dimer),
\begin{eqnarray}
\label{conf}
\left\{\overline{q}^{*}_{\kappa},q^{*}_{\kappa}\right\}
=\left\{\left(\overline{q}^{*}_{0},\sqrt{\frac{4}{3}(mE^*-\overline{q}^{*2}_{0})}\right),\left(\overline{q}^{*2}_{1},\sqrt{\frac{4}{3}(mE^*-\overline{q}^{*}_{1})}\right),\dots,\left(\overline{q}^{*}_{N_{E^*}},\sqrt{\frac{4}{3}(mE^*-\overline{q}^{*2}_{N_{E^*}})}\right)\right\} \ .
\end{eqnarray}
These channels contribute to the quantization condition since $\overline{q}^{*2}_{\kappa}/m<E^*$. This observation makes the analogy to the ``coupled-channel" systems self-evident. For CM energies that are below the dimer energy, $E^*<\overline{q}^{*2}_{\kappa}/m$, the energy is not sufficient to allow the three-particle system to go on-shell. Subsequently, these states can be neglected as they give rise to exponential corrections in volume rather than power-law. Furthermore, similar to the two-body case, the on-shell condition does not fix the directional degrees of freedom of the relative momentum of the $2+1$ system, and therefore it is convenient to upgrade all finite volume quantities into infinite-dimensional matrices in angular momentum.

The calculation of the second term in the expansion of the correlation function, Eq. (\ref{corr3}), is more involved as it comes with one insertion of the three-body Kernel, Fig. (\ref{fig:twoloop}), and due to the one boson exchange contribution couples the momenta running into the loops,
\begin{eqnarray}
C^V_{3,1}(E)&=&
\frac{1}{L^6}\sum_{\textbf{q}_1,\textbf{q}_2}
A_3\left(\mathbf{q}_1\right){i\mathcal{D}^V(E-\frac{{q}^2_1}{2m},|\textbf{P}-\textbf{q}_1|)} \nn\\
\label{Cexd0}
& & \times \left[-ig_{3}-\frac{i{g_2^2}}{E-\frac{\textbf{q}_1^2}{2m}-\frac{\textbf{q}_2^2}{2m}-\frac{(\textbf{P}-\textbf{q}_1-\textbf{q}_2)^2}{2m}+i\epsilon}\right]
{i\mathcal{D}^V(E-\frac{{q}^2_2}{2m},|\textbf{P}-\textbf{q}_2|)}A'_3\left(\mathbf{q}_2\right).
\end{eqnarray}
Although at the first glance, there appears to be poles arising from the exchange boson propagator, one can verify that the poles of the three-body kernel are exactly canceled by the zeros of the full finite volume dimer propagator\footnote{This important observation was first pointed out to us by Michael D\"oring and Akaki Rusetsky for the relativistic three-particle systems \cite{RusetskyDoring}.}. As a result, the only power law volume dependence of such diagrams arise from the poles of the dimer propagator only. Given this observation, it is straightforward to show that
\begin{eqnarray}
C^V_{3,1}(E)&=& 
\int\frac{d^3\textbf{q}_1}{(2\pi)^3}\frac{d^3\textbf{q}_2}{(2\pi)^3}
A_3\left(\mathbf{q}_1\right){i\mathcal{D}^V(E-\frac{{q}^2_1}{2m},|\textbf{P}-\textbf{q}_1|)}
iK_3(\textbf{q}_1,\textbf{q}_2;\mathbf{P},E)
{i\mathcal{D}^V(E-\frac{{q}^2_2}{2m},|\textbf{P}-\textbf{q}_2|)}A'_3\left(\mathbf{q}_2\right)
\nn\\
& - & 2\int\frac{d^3\textbf{q}_1}{(2\pi)^3} \ A_3\left(\mathbf{q}_1\right) {i\mathcal{D}^V(E-\frac{{q}_1^2}{2m},|\textbf{P}-\textbf{q}_1|)} 
 \sum^{N_{E^*}}_{\kappa}\left[K_{3} (\mathbf{q}_1,{q}^*_{\kappa};E^*)\delta\tilde{\mathcal{G}}^V_\kappa({q}_{\kappa}^{*})A'_3(q^*_{\kappa})\right]
\nn\\
\label{Cexd}
& - &
\sum^{N_{E^*}}_{\kappa,\kappa'}
\left[A_3(q^*_{\kappa'})\delta\tilde{\mathcal{G}}^V_{\kappa'}({q}_{\kappa'}^{*})iK_{3} ({q}^*_{\kappa'},{q}^*_{\kappa};E^*)\delta\tilde{\mathcal{G}}^V_\kappa({q}_{\kappa}^{*})A'_3(q^*_{\kappa})\right] \ ,
\end{eqnarray}
where a summation over angular momentum is understood for the terms inside the brackets. The summation over the two-body L\"uscher poles is left explicit.
The result of Eq. (\ref{Cexd}), along with the fact that the dimer propagator can be decomposed in a series over its poles,
\begin{eqnarray}
\label{dimerdecomp}
i\mathcal{D}^{V}(mE-3{q}^2/4,{q})
=\sum^{N_{E^*}}_{\kappa}\frac{iR^V_\kappa}{\overline{q}^{*2}-{\overline{q}}_{\kappa}^{*2}+i\epsilon} \ ,
\label{decomp}
\end{eqnarray}
suggests that the dimer propagator can be upgraded unto a diagonal matrix in the space of $N_{E^*}$ available FV states which is a useful representation when performing the sum over all diagrams contributing to the correlation function. Each element of this matrix is then effectively a single particle propagator with the corresponding FV pole and residue that contain finite volume dependence of the propagators,
 \begin{eqnarray}
 \label{dimerM}
 \left[i\mathcal{D}^{V}(mE-3{q}^2/4,{q})\right]_{\kappa \kappa'}
=\frac{iR^V_\kappa}{\overline{q}^{*2}-{\overline{q}}_{\kappa}^{*2}+i\epsilon} \ \delta _{\kappa \kappa'} \ .
\label{propmatx}
\end{eqnarray}
\begin{figure}[t]
\begin{center}
\subfigure[]{
\includegraphics[scale=0.325]{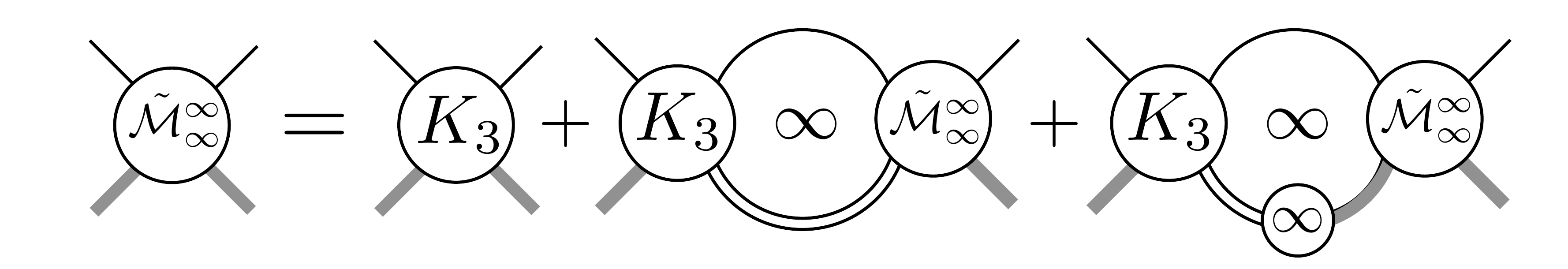}}
\subfigure[]{
\includegraphics[scale=0.325]{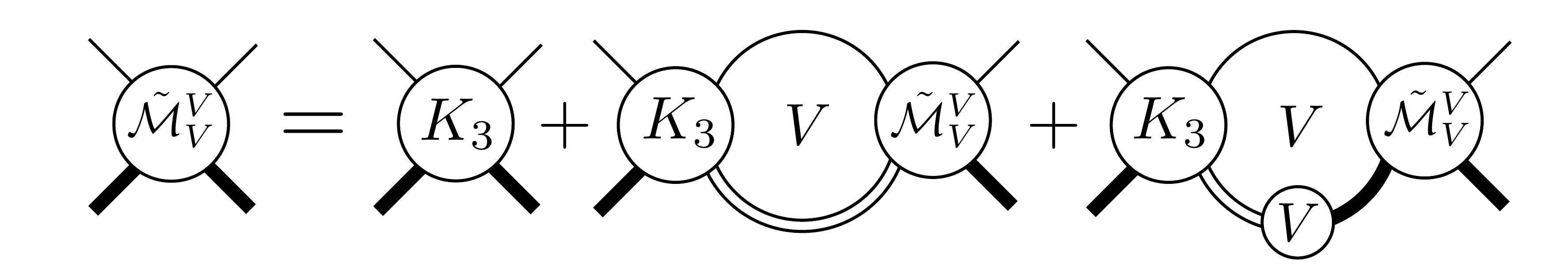}}
\subfigure[]{
\includegraphics[scale=0.325]{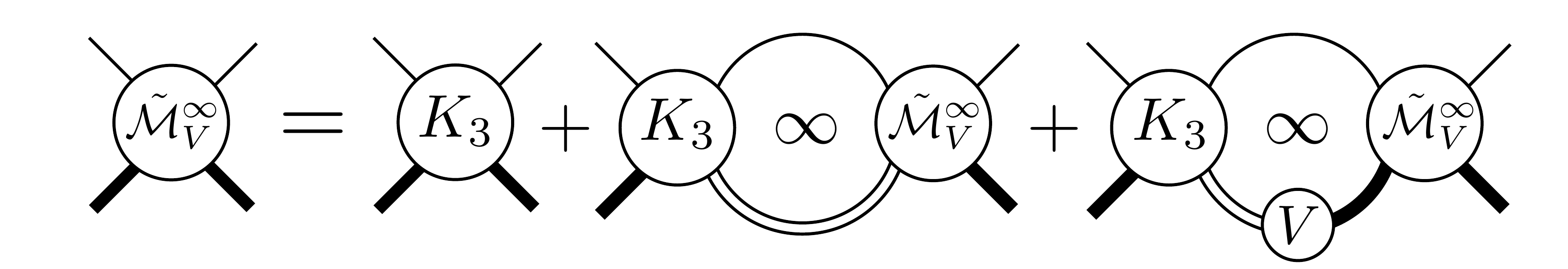}}
\caption{\label{fig:scattamp} a) Diagrammatic representation of the inhomogenous integral equation satisfied by the three-body scattering amplitude in the infinite volume, b) The corresponding sum equation satisfied by the FV scattering amplitude, c) The diagrammatic representation of the integral equation satisfied by $\tilde{\mathcal{M}}^{\infty}_{V}$, Eq. (\ref{STMeq2}).}
\end{center}
\end{figure}

Given the simplifying feature of the FV loop sums as is evident from Eq. (\ref{Cexd}), and the representation of the FV dimer a matrix over available channels, Eq. (\ref{dimerM}), it is straightforward to sum over the infinite number of terms appearing in the boson-dimer correlation function, Eq. (\ref{corr3}). Denoting the boson-FV dimer propagator as $\mathcal{G}^{\infty}_{V}$, one can show that Eq. (\ref{corr3}) is equal to
\begin{eqnarray}
\label{kernelexp}
C^V_3(E,\textbf{P})-C^{\infty}_{3,V}(E,\textbf{P})&=&iA_3\left[(1-\tilde{\mathcal{G}}^{\infty}_{V}\tilde{\mathcal{M}}^{\infty}_{V}) \delta \tilde{\mathcal{G}}^V\frac{1}{1+\tilde{\mathcal{M}}^{\infty}_{V}\delta\tilde{\mathcal{G}}^V}(1-\tilde{\mathcal{M}}^{\infty}_{V}\tilde{\mathcal{G}}^{\infty}_{V})\right]A'_3 \ ,
\end{eqnarray}
where $C^{\infty}_{3,V}(E,\textbf{P}) \equiv iA_3\tilde{\mathcal{G}}^{\infty}_{V}{(1-\tilde{\mathcal{M}}^{\infty}_{V}\tilde{\mathcal{G}}^{\infty}_{V})}A'_3$. $\tilde{\mathcal{M}}^{\infty}_{V}$ is defined as the sum over all infinite volume diagrams containing a boson and a finite volume dimer, Fig. (\ref{fig:scattamp}, c), and can be interpreted as the non-renormalized infinite volume scattering amplitude between a boson and a FV dimer.\footnote{The difference between renormalized and non-renormalized scattering amplitudes will be explained shortly.} That is to say, while the relative momentum between the dimer and boson is continuous in $\tilde{\mathcal{M}}^{\infty}_{V}$, the relative momentum of the two bosons inside the dimer remains discretized,
\begin{eqnarray}
\tilde{\mathcal{M}}^{\infty}_{V}\left(\mathbf{p},\mathbf{k};\mathbf{P},E\right)& = & K_3(\textbf{p},\textbf{k};{\mathbf{P}},E) -
\int\frac{d^3q}{(2\pi)^3}K_3(\textbf{p},\textbf{q};{\mathbf{P}},E){\mathcal{D}^{V}(E-\frac{q^2}{2m},|\mathbf{P}-\mathbf{q}|)} \tilde{\mathcal{M}}^{\infty}_{V}\left(\mathbf{q},\mathbf{k};\mathbf{P},E\right) \nn\\
& = & 
 \tilde{\mathcal{M}}^{\infty}_{\infty}(\textbf{p},\textbf{k};{\mathbf{P}},E) -
\int\frac{d^3q}{(2\pi)^3}\tilde{\mathcal{M}}^{\infty}_{\infty}\textbf{p},\textbf{q};{\mathbf{P}},E)\delta{\mathcal{D}^{V}(E-\frac{q^2}{2m},|\mathbf{P}-\mathbf{q}|)} \tilde{\mathcal{M}}^{\infty}_{V}\left(\mathbf{q},\mathbf{k};\mathbf{P},E\right) \nn\\
\label{STMeq2}
\end{eqnarray}
where $\delta\mathcal{D}^{V}=\mathcal{D}^{V}-\mathcal{D}^{\infty}$ and $\tilde{\mathcal{M}}^{\infty}_{\infty}$ is the non-renormalized infinite volume scattering amplitude, Fig. (\ref{fig:scattamp}, a). For comparison, the full FV scattering amplitude of the dimer-boson system is also depicted in Fig. (\ref{fig:scattamp}, b), where all relative momenta between the three-particles are necessarily discretized.\footnote{One should however note that due to the absence of asymptotic states in a finite volume, the interpretation of this quantity as the scattering amplitude is ambiguous, and it is only introduced as a useful mathematical quantity in analogy with the infinite volume scattering amplitude.} Note that the boson-dimer propagator and $\tilde{\mathcal{M}}^{\infty}_{V}$ are evaluated on-shell where the scattering energies of each boson-dimer channel is given by ${q}^{*2}_{\kappa}/m$.

The poles of the FV correlation function, Eq. (\ref{kernelexp}), determine the spectrum, 
\begin{eqnarray}
\label{QC}
\rm{Det}(1+\tilde{\mathcal{M}}^{\infty}_{V}\delta\tilde{\mathcal{G}}^V)\equiv\rm{det}_{\rm{oc}}\left[\rm{det}_{\rm{pw}}(1+\tilde{\mathcal{M}}^{\infty}_{V}\delta\tilde{\mathcal{G}}^V)\right]=0 \ ,
\end{eqnarray}
where the determinant $\rm{det}_{\rm{oc}}$ is over the $N_{E^*}$ open channels and the determinant $\rm{det}_{\rm{pw}}$ is over the boson-dimer relative angular momentum. In practice it is necessary to perform a truncation over the partial-waves and choose a maximal angular momentum. This quantization condition however incorporate the partial-wave mixing due to the reduced symmetry of the boson-dimer wavefunction in the finite cubic volume as will be discussed in more details in the next section.

The reason that the scattering amplitude quantities introduced above are not renormalized is that unlike single particle operators in a non-relativistic field theory, the dimer field corresponds to an interpolating operator that has overlap with two-particle states, and as a result must be renormalized \cite{Chen:1999tn, Bedaque:1998km}. The renormalization factor in the finite volume can be obtained from the residue of the FV dimer propagator,
\begin{eqnarray}
\label{ZV}
(\mathcal{Z}^{V}_\kappa)^{-1}=i\left.\frac{\partial}{\partial E^*}\frac{1}{i\mathcal{D}^V\left(E-\frac{{q}^2}{2m},|\textbf{P}-\textbf{q}|\right)}\right|_{E^*=\frac{\bar{q}^{*2}_{\kappa}}m+\frac{3q^{*2}_{\kappa}}{4m}}
=\frac{m}{R^V_\kappa} \ .
\end{eqnarray}
Upon renormalizing the dimer field, therefore, one arrives at the normalized scattering amplitudes in the finite volume, e.g. $(\mathcal{M}^{\infty}_{V})_{\kappa \kappa'}=(\mathcal{Z}^{V}_\kappa)^{1/2} (\tilde{\mathcal{M}}^{\infty}_{V})_{\kappa \kappa'} (\mathcal{Z}^{V}_{\kappa'})^{1/2}$.

The quantization condition, Eq. (\ref{QC}), resembles that of the two-body coupled-channel systems as presented in Refs. \cite{Hansen:2012tf, Briceno:2012yi}. As discussed earlier this illustrates that a single on-shell condition does not fix the magnitude of both relative momenta and there is a freedom in scattering in any of \emph{finite} number of available channels. The other characteristic of Eq. (\ref{QC}) is that it does not still provide a algebraic relation between the infinite volume scattering amplitude and the energy eigenvalues of the boson-diboson system, simply because $\tilde{\mathcal{M}}^{\infty}_{V}$ still has possibly large volume corrections arising from FV dimer propagator, Eq. (\ref{STMeq2}). Despite all these complexities, this quantization condition not only gives a better insight into the three-body problem in a finite volume, it automatically reduces to the L\"uscher quantization condition for the bound-state-particle scattering below the bound-state breakup, up to exponential corrections that are due to the size of the bound-state wave-function. This will be discussed in the next section in more details.

\section{Boson-diboson scattering below the breakup threshold \label{recoverL}}
The formalism developed in the previous section does not assume any specific form for the interactions in the three-body system. Therefore the result presented is universal regardless of the nature of the interactions or whether the theory contains any number of two-body or three-body bound-states. In this section, though, we consider a theory with an attractive two-body force which allows a two-body bound-state, a \emph{diboson}. We will show how Eq. (\ref{QC}) reduces to the well-known two-body result below the diboson breakup.

For such energies there is only one state that can go on-shell and introduce power-law volume corrections, the boson-diboson state. By restricting to $l_{\rm max}=0$, the low-energy parametrization of the scattering amplitude becomes that of a two-particle system in an S-wave with masses $m_1=\tfrac{m_2}{2}=m$,
\begin{eqnarray}
\label{Bdscattamp}
{\mathcal{M}}_{Bd}&=&\frac{3\pi}{m}\frac{1}{{q}^*_{0}\cot\delta_{Bd}-i{q}^*_{0}},
\end{eqnarray}
where ${q}_{0}^{*2}/m\equiv \frac{4}{3}(E^*-\bar{q}^{*2}_0/m)$ is the boson-diboson scattering energy in the CM frame, $\bar{q}^{*2}_0/m$ is the boosted diboson FV binding energy, and $\delta_{Bd}$ denotes the boson-diboson scattering phase-shift. However, this is the non-renormalized quantity $\tilde{\mathcal{M}}^{\infty}_{V}$ that appears in the QC, Eq. (\ref{QC}), and not the physical scattering amplitude. Here we argue that by introducing a systematic error of the order of $e^{-\gamma_d L}/L$ to the final result, Eq. (\ref{dbQC}), the scattering phase shifts can be derived from the QC, Eq. (\ref{QC}) after replacing $\delta\tilde{\mathcal{G}}^V\tilde{\mathcal{M}}^{\infty}_{V}$ with $\delta\mathcal{G}^V\mathcal{M}^{\infty}_{\infty}\equiv\delta\mathcal{G}^V{\mathcal{M}}_{Bd}$. $\gamma_d$ denotes the infinite volume binding momentum of the diboson which satisfies,
\begin{eqnarray}
\label{2QCI}
\left. \left(\overline{q}^{*}\cot{\delta_d}-i\overline{q}^{*}\right)\right|_{\overline{q}^{*}=i\gamma_d}=0 \ .
\end{eqnarray}

The first step to prove this claim is to note that the bound-state pole of the FV dimer propagator is exponentially close to the bound-state pole of the infinite volume dimer, $\bar{q}^{*}_0=i\gamma_d+\mathcal{O}(e^{-\gamma_d L}/L)$,
which is evident from Eq. (\ref{2QCV}) after analytically continuing the momentum $\tilde{q}^*_0$ to the imaginary axis. Now in evaluating $\tilde{\mathcal{M}}^{\infty}_{V}$, one needs to perform a series of coupled integrals as is given in the first line of Eq. (\ref{STMeq2}). For negatives energies, the only singularity of the integrands in the range of integration occurs when the diboson pole of the FV dimer propagators is reached. The contribution to the integrals due to this singularity is proportional to the residue of the FV dimer at that pole. Since the residue of the infinite volume dimer propagator at the diboson pole,
\begin{eqnarray}
\label{residuedef}
R^{\infty}_{d}=-\frac{mr}{2} \left[\frac{\partial}{\partial \overline{q}^{*2}}\left(\overline{q}^{*}\cot\delta_d-i\overline{q}^{*}\right)|_{\overline{q}^{*2}=-\gamma_d^2}\right]^{-1} \ ,
\end{eqnarray}
is exponentially close to its FV counterpart, Eq. (\ref{residuedef}),\footnote{There is another correction to the residue function at the diboson pole that occurs at $\mathcal{O}\left(e^{-\gamma_d L} / \gamma_d L\right)$. Since for $\gamma_dL>1$ the diboson does not fit in the volume, and the finite volume formalism is no longer valid, one must make sure to use sufficiently large volumes for shallow bound-states so that $\gamma_dL \gg 1$. It then follows that these corrections are subleading compared to the $\mathcal{O}(e^{-\gamma_d L})$ corrections in Eq. (\ref{RVRI}) and could be ignored.}
\begin{eqnarray}
\label{RVRI}
R^{V}_{d}=R^{\infty}_{d}\left[1+\mathcal{O}(e^{-\gamma_d L})\right] \ ,
\end{eqnarray}
one can replace $\mathcal{D}^V$ with $\mathcal{D}^{\infty}$ up to the exponential corrections that scale by the size of the bound-state wave-function. Consequently, from Eq. (\ref{STMeq2}) one observes that $\tilde{\mathcal{M}}^{\infty}_{\infty}$ is equal to $\tilde{\mathcal{M}}^{\infty}_{V}$ up to exponentially small corrections. Note that $\tilde{\mathcal{M}}^{\infty}_{V}$ and $\tilde{\mathcal{M}}^{\infty}_{\infty}$ are renormalized differently, however, the finite volume dimer field renormalization factor $\mathcal{Z}^V$, Eq. (\ref{ZV}) is exponentially close to the renormalization factor of the infinite volume diboson field $\mathcal{Z}^{\infty}$ around the bound-state pole, which is defined as
\begin{eqnarray}
\label{ZI}
(\mathcal{Z}^{\infty}_d)^{-1}=i\left.\frac{\partial}{\partial E^*}\frac{1}{i\mathcal{D}^{\infty}\left(E-\frac{{q}^2}{2m},|\textbf{P}-\textbf{q}|\right)}\right|_{E^*=-\frac{\gamma_d^2}m+\frac{3q^{*2}_{d}}{4m}}
=\frac{m}{R^{\infty}_d} \ .
\end{eqnarray}
Therefore one can approximate $\delta\tilde{\mathcal{G}}^V\tilde{\mathcal{M}}^{\infty}_{V}=\delta\mathcal{G}^V\mathcal{M}^{\infty}_{V}$ in Eq. (\ref{QC}), with $\delta\mathcal{G}^V\mathcal{M}^{\infty}_{\infty}$ for elastic processes that occur in this energy regime up to these exponential corrections as stated.

Keeping in mind these exponential corrections, one can now apply the expression for the scattering amplitude in Eq. (\ref{Bdscattamp}), to Eq. (\ref{QC}). Using Eqs. (\ref{dgv}), (\ref{gtilde}), one can recover the two-particle quantization condition for two-particle systems up to the exponential corrections explained above as shown in Eq. (\ref{dbQC}) and reiterated here for clarity
\begin{eqnarray}
\label{dbQC2}
{q}^*_{0}\cot\delta_{Bd} &=&4\pi \ c^P_{00}({q}_{0}^{*})+\eta\frac{e^{-\gamma_d L}}{L}.
\end{eqnarray}
This result confirms the postulation and numerical verification made by Bour \emph{et al.}~\cite{Bour:2012hn} that upon subtracting off the FV binding energy of the bound-state from the total energy of the three-particle system, the scattering energy eigenvalues of the bound-state-particle system can be reliably related to the scattering phase shift of the system through the use of L\"uscher formula for two-body systems after extrapolating to the infinite volume limit. This offers the practitioner a reliable method to extract the infinite volume phase shift of elastic bound-state-particle scattering by fitting to an exponential form.

This result also illustrates that in order to obtain the boson-diboson scattering phase shift, not only one needs to determine the boosted three-particle energy spectrum, but also needs to obtain the scattering parameters of the boosted two-particle system first. It is also evident that if the interactions support a boson-diboson bound-state, a \emph{triboson}, after analytically continuing the momentum in Eq. (\ref{dbQC}) to the imaginary plane, ${q}^*_{0}=i\gamma_{Bd}$, the binding energy of the three-particle system, $B_3=\frac{3\gamma_{Bd} ^2}{4m}$, can be obtained easily via Eq. (\ref{dbQC}), as is well-known for bound-states appearing in the two-body sector~\cite{luscher1, luscher2, pds3}. Alternatively, one can also solve for the triboson poles of the FV scattering amplitude from the FV counterpart of the STM equation, Fig. (\ref{fig:scattamp}, b), as is pursued in Refs.~\cite{Kreuzer:2008bi, Kreuzer:2009jp, Kreuzer:2010ti, Kreuzer:2012sr}. 

The boson-diboson QC, Eq. (\ref{dbQC}), is a low-energy approximation of Eq. (\ref{QC}), which at NLO has two sources of exponential corrections. First, the QC receive corrections associated with the finite volume binding momentum of the diboson which scale like $\mathcal{O}(e^{-\sqrt{2}\gamma_d L}/L)$ at next to leading order. It also acquires exponential corrections associated with the truncation of off-shell states appearing in the decomposition of the dimer propagator, Eq. (\ref{dimerdecomp}), as mentioned before. More explicitly, the next excited state of the boson-diboson system corresponds to a CM scattering energy of ${q}_{1}^{*2}/m\equiv\frac{4}{3}(E^*-\bar{q}^{*2}_1/m)$, where $\bar{q}^{*}_1$ is the boosted momentum for an unbound two-boson system. For $E^*<\bar{q}^{*2}_1$/m, the three-boson scattering energy, ${q}_{1}^{*2}/m$, is negative which leads to exponential corrections of $\mathcal{O}\left(e^{-{{q}_{1}^{*}}L}/L\right)$ to the single-channel QC, Eq. (\ref{dbQC}), which however are subleading compared to $\mathcal{O}\left(e^{-{{q}_{0}^{*}}L}/L\right)\sim\mathcal{O}(e^{-\gamma_d L}/L)$ corrections. For sufficiently high energies, these exponential corrections become power-law in the volume, and one necessarily has to study a coupled-channel system made up of a boson-diboson state and a three-boson state. For energies just above the diboson breakup Eq. (\ref{QC}) can be written as  
\begin{eqnarray}
\label{QCbreakup}
\left(1+\tilde{\mathcal{M}}^{\infty}_{V,Bd-Bd}~\delta\tilde{\mathcal{G}}^V_{Bd}\right)\left(1+\tilde{\mathcal{M}}^{\infty}_{V,BBB-BBB}~\delta\tilde{\mathcal{G}}^V_{BBB}\right)=|\tilde{\mathcal{M}}^{\infty}_{V,Bd-BBB}|^2~\delta\tilde{\mathcal{G}}^V_{Bd}~\delta\tilde{\mathcal{G}}^V_{BBB}\ ,
\end{eqnarray}
where $\delta\tilde{\mathcal{G}}^V_{Bd}$ and $\delta\tilde{\mathcal{G}}^V_{BBB}$ are respectively the boson-diboson and three boson propagators, $\tilde{\mathcal{M}}^{\infty}_{V,\kappa-\kappa'}$ denotes the elements of $\tilde{\mathcal{M}}^{\infty}_{V}$ for the $\kappa^{th}$ ($\kappa'^{th}$) initial (final) state. For such energies, the approximations made before are no longer valid and determination of infinite volume scattering cross sections from the finite volume spectrum requires numerically solving an integral equation for $\tilde{\mathcal{M}}^{\infty}_{V}$, Eq. (\ref{STMeq2}).

Lastly we comment on the systematic uncertainties associated with the dimer formalism and partial-wave mixing. Assume, for example, that both the dimer and the boson-dimer wavefunctions are projected onto the $A_1^+$ irreducible representation of the cubic group and that the three particles are degenerate. Then in the boson-dimer CM frame, the system has an overlap with $(J_{d},J_{Bd})=(0,0)$ as well as $(J_{d},J_{Bd})=\{(2,0),(4,0),(0,4),(2,4),(2,6),\ldots\}$ angular momentum states, with the leading contamination arising from the D-wave dimer. As discussed in Sec. \ref{3body}, this is due to the the fact that the dimer in this $2+1$ body set-up is boosted and its symmetry group in its CM frame is reduced compared to the original cubic group \cite{movingframe}. If one then proceeds to consider a reference frame where the dimer-boson system has non-zero momentum, then the ground state will have overlap with $(J_{d},J_{Bd})=(0,0)$ as well as $(J_{d},J_{Bd})=\{(0,1),(2,0),(2,1),(0,2),(2,2),\ldots\}$ angular momentum states. This is because the dimer-boson is effectively a two-particle system where one of the particles is twice as massive as the other, and therefore S and P-wave mixing is unavoidable \cite{Fu:2011xz}. As a result one needs to simultaneously determine S and P-wave scattering parameters. Note that although the dimer field used in this paper is an S-wave filed which does not lead to inclusion of higher partial-waves in the two-body QC, the boson-diboson scattering QC, Eq. (\ref{dbQC}) fully incorporates the partial-wave mixing in the space of the boson-diboson angular momentum states.

\subsection{On the quantization conditions derived by Guo, \textit{et al.} in the isobar approximation \label{isobarf}}

Developing a finite volume multi-particle formalism not only is crucial in studying nuclear reactions from LQCD, but also is a necessary step towards reliable LQCD studies of higher excited states as well as hadronic resonances. With this latter goal in mind, authors of Ref. \cite{Guo:2012hv} have attempted recently to present such a FV formalism for three-body hadronic processes that have a dominant two-particle resonant channel. Such a study is motivated since, as an example, the decay of $J/\psi$ to three pions has been shown to be well-approximated by decaying to a pion as well as the $\rho$ resonance where the $\rho$ can subsequently decay into two pions \cite{Bai:2004jn}. Modeling a process that involves a three-particle final state via such a quasi two-particle intermediate state is called the isobar approximation \cite{Goradia:1975ec, Ascoli:1975mn, Hwa:1963, Aaron:1973ca}, and has been used by Guo \textit{et al.} to study the spectrum of the three-particle system in the finite volume. Here we argue that the result presented by these authors fails to describe the three-particle processes that can be studied via the isobar model in the infinite volume, and by comparing their quantization condition with that of presented in this paper, confirm the validity of their result in one particular case that is different from processes which the isobar model concerns.

Before discussing the result of Ref. \cite{Guo:2012hv}, it is necessary to briefly review some important features of a resonance that makes it challenging to be extracted via a LQCD calculation. As is well known, a resonance is not an isolated eigenstate of the hamiltonian but rather a resonance in multi-particle scattering channels that have the same quantum numbers as the resonance state. In the limiting case where the resonance has an infinitely narrow width, one however recovers an isolated state of the theory. The key is that the resonance itself should not be confused with a composite object, e.g. the $\rho$ meson can not be interpreted as a two-pion state; as soon as the dynamical coupling of $\rho$ to the two-pion state is turned off, it no longer can decay to two pions. That is a key difference between a resonance that dominates the two-particle scattering amplitude, and a two-particle bound-state. As a result, the extraction of two-body bound-states follows directly from the identification of the negative \emph{interaction} energy levels of the two-particle spectrum, and its binding momentum in the finite volume is directly related to the scattering phase shift via the standard two-body L\"uscher formula. To extract a resonance state from a LQCD calculation, however, requires mapping out the scattering phase shifts of the corresponding two-particle scattering channel as a function of energy using the L\"uscher formalism (for a recent implementation of this technique see Ref. \cite{Dudek:2012xn}). In addition, since most of resonances lie above the inelastic threshold, their extractions require studying a coupled-channel multi-particle scattering system.

With this difference in mind, let us discuss the result derived by Guo, \textit{et al.} for three-particle systems in a finite volume within the isobar approximation. Based on a Hamiltonian formalism approach, Guo \textit{et al.} presented two sets of L\"uscher formula that are derived from a relativistic Lippmann-Schwinger equation. One of these formulae describes the scattering of the two particles that compose the isobar in terms of the invariant mass of the isobar in the $2+1$ system. This equation corresponds to Eq. (\ref{2QCV}) of this article where the scattering phase shift of the two-particle sub-system is related to the relative momentum of the particles in this sub-system. The other equation however, is effectively another two-particle L\"uscher formula describing the scattering of an isobar off the spectator particle. This relation offers an algebraic relation between the phase shift of the particle-isobar scattering system in terms of the relative momentum of the isobar and particle in the finite volume. Note that the relative momentum of the isobar and particle is correlated with the solutions to the L\"uscher equation for the isobar system at any given total energy which is a direct consequence of three-particle kinematics.

There are two features of these quantization conditions that make them unrealistic solutions to the three-body problem within the model considered. First of all, as is discussed in detail through this article, for any given CM energy of the three-particle system, there are more than one configuration of momenta in the finite volume, see Eq. (\ref{conf}), that can go on-shell, and the corresponding channels need to be included in the quantization condition simultaneously. Although the authors of Ref. \cite{Guo:2012hv} point out that there are multiple combinations of relative momenta for each given energy that could satisfy their quantization conditions, their results fail to incorporate this coupled-channel feature of the problem. As a direct consequence, not only systematic errors due to nearby off-shell states of the $2+1$ system can not be addressed by their derivation, but most importantly their quantization conditions fail to describe the true behavior of the system as soon as an extra on-shell state is available at a given energy. It is important to reiterate that these ``states" are nothing more than the kinematically allowed configurations of the three-particle system. 

The other misleading feature of these quantization conditions is that although the formalism is presented in the framework of the isobar model, no connection to the resonant nature of the two-particle sub-system has been made. Most importantly, the isobar, although introduced as a sharp resonance, has been treated as a composite system of two-particles. This composite system then has been studied via a two-body L\"uscher formula in a finite volume similar to two-body bound-states. The question that arises is that if the resonance is being treated as an isolated particle that scatters off another particle in this setup, which is a reasonable assumption in the limit of infinitesimal width, how could its energy spectrum be related to the scattering parameters of the two-body sub-system? On the other hand, as soon as a dynamical coupling to the two-particle state is turned on (which is the case for systems with three-body final states), the resonance is no longer an isolated state of the hamiltonian and its energy is not given by the standard L\"uscher formula, in contrary to the two-body bound and scattering states. One needs to take into account this distinction for one primary reason. A resonance state can get large volume corrections in a finite volume, and as a result, the correspondence between the isobar-particle spectrum in the finite volume and the isobar-particle scattering amplitude through a simple L\"uscher formula is ambiguous. This is analogous to our discussion of the $2+1$ systems above the breakup where the FV two-particle state receives large power-law volume corrections and the boson-FV dimer scattering amplitude can no longer be approximated by the three-particle infinite volume scattering amplitude. That being said, our formalism demonstrates how one can recover the infinite volume scattering amplitude from the spectrum, but by no means should one expect an algebraic L\"uscher quantization condition for such a quasi $2+1$ system.

There is however one limiting case where the result of Ref. \cite{Guo:2012hv} is valid. As was discussed in Sec. \ref{recoverL}, the scattering phase shifts of boson-diboson scattering can be extracted from an algebraic L\"uscher quantization condition, up to exponential corrections with the leading piece scale by $e^{-\gamma_d L}/L$, Eq. (\ref{dbQC}). This is a valid description of the process since in the energy regime where the two-body bound-state can not break up, there is only one on-shell momentum configuration that could lead to power-law corrections in volume. The quantization conditions presented by Guo \textit{et al.} is therefore valid for theories that allow a deeply two-body bound-state, in order for the exponential corrections to be negligible, and in a energy regime where the bound-state can not disintegrate into its constituent particles. Then effectively one has an elastic scattering between two particles that can be studies via a single-channel L\"uscher quantization condition, where the energy of one of the particles (the bound-state) is determined from another L\"uscher formula. It is only for these particular conditions that the equations given in Ref. \cite{Guo:2012hv} are valid. So in conclusion, although the quantization conditions of Ref. \cite{Guo:2012hv} could be applicable to deeply bound-state particle scattering below the breakup threshold, it cannot make a valid connection between the scattering parameters of an isobar model in infinite volume and the energy eigenvalues of three particles in a finite volume.

\section{Summary and Conclusion \label{summ}}
Determining nuclear reaction cross sections directly from the underlying theory of QCD will impact our understanding of a wide range of phenomena. Currently LQCD is the only reliable way to carry out such ambitious program provided that the formalism that connects the physical scattering amplitudes to the lattice multi-nucleon correlation functions is put in place. With this in mind, we have determined a model-independent representation of the quantization condition for energy eigenvalues of three identical bosons in a finite volume with the periodic boundary conditions. Using a non-relativistic EFT, the FV three-particle spectrum has been shown to be related to the infinite volume S-matrix elements. For arbitrary energies, this correspondence requires solving an integral equation. With nuclear systems in mind, close attention is paid to scalar theories that support a two-body bound-state. It is shown that for energies below the diboson breakup, the quantization condition reduces to the well known L\"uscher result for two particles with unequal masses, with exponential corrections dictated by the size of the diboson, Eq. (\ref{dbQC}). Although physically intuitive, this is a non-trivial observation that will require the Lattice practitioner to extrapolate the phase shifts obtained from the L\"uscher formula for the particle bound-state elastic scattering to the infinite volume.

In deriving the quantization condition for the three-boson system, an auxiliary dimer field can be used for convenience. This dimer field has the same quantum numbers as of the two-boson system in an S-wave, and enables one to study the three-body system as an effective 2+1 system. This simplification comes at the cost of neglecting the partial-wave mixing associated with a cubic finite volume with the periodic boundary conditions. By evaluating poles of the full three-boson correlation function in the finite volume, the quantization condition for the three-particle system with non-zero total momentum is derived. The poles are given by a determinant condition, Eq. (\ref{QC}), where the determinant is taken over dimer-boson relative angular momentum states as well as $N_{E^*}$ available boson-dimer eigenstates for each CM energy, $E^*$. As is shown, the corresponding quantization condition has strong parallels with two-body coupled-channel systems \cite{Hansen:2012tf, Briceno:2012yi}. However one has to be careful that this is not the physical scattering amplitude that directly shows up in the QC, but rather the scattering amplitude of a boson-FV dimer system. These two quantities are related to each other through an integral equation, Eq. (\ref{STMeq2}). {This manifestly shows that the determination of the scattering quantities of the three-particle system from the energy eigenvalues does not simply follow from two coupled L\"uscher formulae for two-particle sub-system, as well as 2+1 scattering system, as suggested in Ref. \cite{Guo:2012hv}.} It has been demonstrated that as the energy of the system increases, more three-particle states can go on-shell, and as a result large FV correction due to the size of the two-particle sub-system can not be negleted. This feature is clearly embedded in the QC presented, Eqs. (\ref{QC}), (\ref{STMeq2}).

Furthermore, as is explained in detail, the exponential volume corrections from the off-shell excited states of the dimer are accounted for in the full quantization condition. For sufficiently high energies, these exponential corrections become power-law in volume and can no longer be neglected. Then one would have to consider a coupled-channel system where the number of channels are determined by the total CM energy of the three-particle system, as shown in Eq. (\ref{QCbreakup}) for energies just above the diboson breakup. The formalism presented considers three-particle with non-zero total momentum which eventually allows for more independent measurements at a given energy. This however leads to a practical complexity as the symmetry of the system is reduced, and the ground state of the system is expected to mix with the P-wave scattering state \cite{Bour:2011ef, Davoudi,Fu:2011xz}. The quantization condition derived predicts this mixing between S and P partial-waves, and indicates that the truncation of the determinant condition at S-wave could, in practice, introduce large systematics to the calculation.

With these observations at hand, we argue that future LQCD studies of nuclear reactions and resonances involving three-particle states will require the following steps. First, one needs to reliably determine scattering phase shifts for the two-particle sector from which one can obtain the boosted L\"uscher poles as a function of the boost momenta and energy. From there, one would proceed to obtained the three-particle spectrum. This requires high statistics to obtain multiple states with clean signals. Also in order to disentangle the coupled-channel nature of the three particle system, these calculations need to be performed with different boosts and in different volumes. In addition one has to simultaneously determine energy eigenvalues of three-particle states in different irreps of the cubic group to correctly deal with the partial wave mixing which is more severe than the two-particle sector. All of thes information should be simultaneously fitted to numerically solve the quantization condition presented. This would lead to an accurate determination of the three-body Bethe-Salpeter Kernel, which encodes all of the infinite volume physics up to the four-particle inelastic threshold.

\subsection*{Acknowledgment}

We would like to thank Martin J. Savage for motivating efforts to study nuclear reactions using lattice QCD. We also thank him for many fruitful discussions and for his feedback on the first manuscript of this paper. We also would like to thank Maxwell T. Hansen, Michael D\"oring, Akaki Rusetsky, Paulo F. Bedaque, Alan O. Jaminson, Stephen R. Sharpe, David B. Kaplan and Thomas C. Luu for useful conversations. RB and ZD were supported
in part by the DOE grant DE-FG02-97ER41014.

\bibliography{bibi}
\end{document}